\documentstyle[preprint,aps]{revtex}
\begin{document}
\preprint{hep-th/0106047}
\title{ Superluminal Noncommutative Photons }

\author{Rong-Gen Cai\footnote{email address: cai@het.phys.sci.osaka-u.ac.jp}}
\address{Institute of Theoretical Physics, Chinese Academy of Sciences,
  \\
    P.O. Box 2735, Beijing 100080, China \\ 
 Department of Physics, Osaka University, Toyonaka, Osaka 560-0043, Japan} 
\maketitle
\begin{abstract}
With the help of the Seiberg-Witten map, one can obtain an effective action of 
a deformed QED from a noncommutative QED. Starting from the deformed QED, we 
investigate the propagation of photons in the background of electromagnetic 
field, up to the leading order of the noncommutativity parameter. In our 
setting (both the electric and magnetic fields are parallel to the coordinate 
axis $x^1$ and the nonvanishing component of the noncommutativity parameter is
$\theta^{23}$), we find  that the electric field has no effect on the 
propagation of photons, but the velocity of photons can be larger than the
speed of light ($c=1$) when the propagating direction of photons is 
perpendicular to the direction of background magnetic field, while the 
light-cone condition does not change when the propagating direction is parallel 
to the background magnetic field. The causality associated with the 
superluminal photons is discussed briefly. 
\end{abstract}

\newpage

Recently much attention has been focused on the noncommutativity of 
spacetime and its consequences in various aspects. Many interesting  
phenomena have been found. For example, there exists an intriguing UV/IR 
mixing~\cite{Minw} in the perturbative dynamics of noncommutative field 
theories (field theories in a noncommutative spacetime). The degrees of 
freedom are reduced drastically in the non-planar sector of noncommutative
field theories~\cite{Biga,CO1,Fisc}. The energy levels of the hydrogen atom 
and
the Lamb shift in the noncommutative quantum electrodynamics (QED) have 
been found to deviate from those in the usual QED both on the classical and
quantum levels, from which a constraint on the noncommutativity parameter
of spacetime can be obtained by comparing with experimental data~\cite{CST}.
The noncommutativity of spacetime has also significant effects on the 
cosmology of the early universe. For instance, it has been
shown that the noncommutativity in spatial coordinates can generate magnetic
field in the early universe on a horizon scale, and due to its dependence
of the number density of massive charged particles, one can trace back the
temperature dependence of the noncommutativity scale from the bounds on 
the primordial magnetic field coming from nucleosynthesis~\cite{MS}. Further,
it was argued that due to the noncommutativity of spacetime, the inflation
induced fluctuations become non-Gaussian and anisotropic, and the short
distance dispersion relations are modified~\cite{Chu}.

By definition, the coordinates in a noncommutative spacetime satisfy
\begin{equation}
\label{eq1}
[x^{\mu},x^{\nu}]=i \theta^{\mu\nu},
\end{equation}
where $\theta^{\mu\nu}$ is an antisymmetric constant matrix and is of 
dimension of $ (length)^2$. The action of a noncommutative field theory can 
be constructed by replacing the ordinary products between field operators by
the $\star$-products in the action of the corresponding ordinary field theory. 
The $\star$-product
is defined as follows,
\begin{equation}
\label{eq2}
f\star g (x)=e^{\frac{1}{2}\theta^{\mu\nu}\partial_{x^{\mu}}
   \partial_{y^{\nu}}}f(x)g(y)|_{y=x}.
\end{equation}
The Lagrangian of the photon sector of the noncommutative QED is~\cite{Haya} 
\begin{equation}
\label{eq3}
{\cal L}_1 =-\frac{1}{4}\hat F_{\mu\nu}\star \hat F^{\mu\nu},
\end{equation}
where $\hat F_{\mu\nu} =\partial_{\mu} \hat A_{\nu}-\partial_{\nu}
\hat A_{\mu}-i[\hat A_{\mu},\hat A_{\nu}]_{\star}$. Usually one may start with 
the action of a certain noncommutative field theory, calculate some physical
quantities and then compare with experimental data. However, 
it is well-known that there is a so-called Seiberg-Witten map~\cite{SW} 
between a noncommutative gauge field and an ordinary gauge field. The 
physical equivalence is guaranteed between the noncommutative gauge field and 
the ordinary gauge field in the mapping. The Seiberg-Witten map is~\cite{SW}
\begin{equation}
\label{eq4}
\hat A_{\mu}=A_{\mu}-\frac{1}{2}\theta^{\alpha\beta}A_{\alpha}
  (\partial_{\beta}A_{\mu}+F_{\beta\mu}),
\end{equation}
to the leading order of $\theta^{\mu\nu}$. Substituting this into (\ref{eq3})
and using the $\star$-product (\ref{eq2}), one can obtain~\cite{GRS}
\begin{equation}
\label{eq5}
{\cal L}_2 = -\frac{1}{4}F_{\mu\nu}F^{\mu\nu}-\frac{1}{2}\theta^{\alpha\beta}
   F_{\alpha\mu}F_{\beta\nu}F^{\mu\nu} +\frac{1}{8}\theta^{\alpha\beta}
  F_{\alpha\beta}F_{\mu\nu}F^{\mu\nu},
\end{equation}
to the leading order of $\theta$. This Lagrangian can be regarded as the one
of the photon sector of the ordinary QED with some correction terms associated
with the noncommutativity of spacetime. Physically, the two actions 
(\ref{eq3}) and (\ref{eq5}) are equivalent to each other, up to the leading 
order of $\theta$. Therefore the action (\ref{eq5}) plus fermion part could be 
a good starting point to give some constraints on the realistic noncommutative
field theories~\cite{Bich,Carr}.

The form of action (\ref{eq5}) is reminiscent
of the one-loop effective action of ordinary QED, which is 
\begin{equation}
\label{eq6}
{\cal L}_3=-\frac{1}{4}F_{\mu\nu}F^{\mu\nu}-\frac{\alpha^2}{36 m^4_e}
   (F_{\mu\nu}F^{\mu\nu})^2 +\frac{7\alpha^2}{90m^4_e}F_{\mu\nu}F_{\sigma\tau}
   F^{\mu\sigma}F^{\nu\tau},
\end{equation}
where $m_e$ is the mass of electron,  $\alpha$ is the fine structure 
constant, and last two terms are the Euler-Heisenberg ones. Taking into 
account the one-loop correction, thirty years ago, Adler~\cite{Adler} 
 showed that the velocity of photons propagating in an anisotropic vacuum
given by an external constant uniform magnetic field ${\rm \bf B}$, is 
shifted and depends on the direction of polarizations, 
\begin{equation}
\label{eq7}
v_{\parallel}=1 -\frac{8}{45}\alpha^2 \frac{{\rm \bf B}^2}{m^4_e}\sin^2\phi,
\ \ \ 
v_{\perp}=1-\frac{14}{45}\alpha^2\frac{{\rm \bf B}^2}{m^4_e}\sin^2\phi,
\end{equation} 
where $\phi$ is angle between the magnetic field ${\rm \bf B}$ and the 
direction of the propagation.  This is just the electromagnetic    
birefringence phenomenon. Note that here the velocity of photons is always 
smaller than the speed of light $(c=1)$.

With this in mind, it would be of interest to investigate the propagation
of photons within the effective action (\ref{eq5}) in the
background of electromagnetic field. This is just the aim of this note. 
Varying the action (\ref{eq5}) yields the equation of motion
\begin{eqnarray}
\label{eq8}
&& \partial_{\mu} F^{\mu\nu} +\partial _{\mu}(\theta^{\alpha\beta}F_{\alpha}
  ^{\ \mu}F_{\beta}^{\ \nu}) -\partial _{\mu}(\theta^{\alpha \nu}
   F_{\alpha\lambda}F^{\lambda\mu}) \nonumber \\
&&~~~~~~~~-\frac{1}{2}\partial_{\mu}(\theta^{\alpha\beta}F_{\alpha\beta}
      F^{\mu\nu}) +\partial_{\mu}(\theta^{\alpha\mu}F_{\alpha\lambda}
     F^{\lambda\nu}) -\frac{1}{4}\partial_{\mu}(\theta^{\mu\nu}
     F_{\delta\lambda}F^{\delta\lambda})=0.
\end{eqnarray}
In order to study the propagation of photons, we expand the field strength
using background field method,
\begin{equation}
\label{eq9}
F^{\mu\nu}=\bar F^{\mu\nu} +\tilde f^{\mu\nu},
\end{equation} 
where $\bar F_{\mu\nu}$ is the background electromagnetic field. Substituting
this into (\ref{eq8}) and linearizing the equation in $\tilde f^{\mu\nu}$, 
we obtain the equation describing the propagation of photons
\begin{eqnarray}
\label{eq10}
&& \partial_{\mu} \tilde f^{\mu\nu} +\theta^{\alpha\beta}\bar F_{\alpha}^{\
 \mu}\partial_{\mu}\tilde f_{\beta}^{\ \nu} -\theta^{\alpha\nu}\bar F^{\lambda
   \mu}\partial_{\mu}\tilde f_{\alpha \lambda}
   -\frac{1}{2}\theta^{\alpha\beta}\bar F^{\mu\nu}\partial_{\mu}
   \tilde f_{\alpha\beta}  \nonumber \\
&&~~~~~~~ +\theta^{\alpha\mu}\bar F_{\alpha\lambda}\partial_{\mu}\tilde 
    f^{\lambda \nu} +\theta^{\alpha \mu}\bar F^{\lambda \nu}\partial_{\mu}
    \tilde f_{\alpha \lambda}
   -\frac{1}{2}\theta^{\mu\nu}\bar F_{\delta \lambda}\partial_{\mu}
    \tilde f^{\delta\lambda}=0,
\end{eqnarray}
where the background field has been assumed to be a constant electromagnetic
field (namely, $\bar F_{\mu\nu}$ is assumed to be a constant).  In deriving
(\ref{eq10}),  we dropped some terms like  $\theta$ times
$\partial_{\mu} \tilde f^{\mu \lambda}$ (these are of the order $\theta^2$) 
since we keep terms up to the leading order of $\theta$ only. 

In order to investigate the propagation of photons, a simple method is to
employ the geometric optic approximation~\cite{DH}, in which one can 
write
\begin{equation}
\label{eq11}
\tilde f_{\mu\nu} =f_{\mu\nu}e^{i\omega},
\end{equation}
where $f_{\mu\nu}$ is a slowly varying amplitude and $\omega$ is the rapidly
varying phase. The wave vector is $k_{\mu}=\partial_{\mu}\omega$. In quantum
mechanics, it can be viewed as the momentum of the photons. Thus, from
the Bianchi identity,
$\partial _{\lambda}F_{\mu\nu} +\partial_{\mu}F_{\nu\lambda}
  +\partial_{\nu}F_{\lambda\mu}=0$,
one has $k_{\lambda}f_{\mu\nu}+k_{\mu}f_{\nu\lambda} +k_{\nu}f_{\lambda\mu}
=0$. Furthermore, we can write 
\begin{equation}
\label{eq12}
f_{\mu\nu}=k_{\mu}a_{\nu}-k_{\nu}a_{\mu},
\end{equation}
where the vector $a_{\mu}$ can be interpreted as the polarization vector
of the photons and satisfies $k_{\mu}a^{\mu}=0$. In this case, the equation
(\ref{eq10}) reduces to
\begin{eqnarray}
\label{eq13}
&& k^2 a^{\nu} +\theta^{\alpha\beta}\bar F_{\alpha}^{\ \mu}k_{\mu}
  (k_{\beta}a^{\nu} -k^{\nu}a_{\beta}) -\theta^{\alpha \nu}\bar F^{\lambda\mu}
  k_{\mu}(k_{\alpha}a_{\lambda}-k_{\lambda}a_{\alpha})
  \nonumber \\
&&~~~~~~-\frac{1}{2}\theta^{\alpha\beta}\bar F^{\mu\nu}k_{\mu}
   (k_{\alpha}a_{\beta}-k_{\beta}a_{\alpha}) 
  +\theta^{\alpha\mu}\bar F_{\alpha \lambda}k_{\mu}(k^{\lambda}a^{\nu}
   -k^{\nu}a^{\lambda}) \nonumber \\
&&~~~~~~~ +\theta^{\alpha\mu}\bar F^{\lambda \nu}k_{\mu}(k_{\alpha}a_{\lambda}
   -k_{\lambda}a_{\alpha}) -\frac{1}{2}\theta^{\mu\nu}
   \bar F_{\delta\lambda}k_{\mu}(k^{\delta}a^{\lambda}
   -k^{\lambda}a^{\delta})=0. 
\end{eqnarray}  
Now it is a position to choose an appropriate background. It has been already 
shown that a field theory with only nonvanishing space-time component of 
noncommutative matrix, $\theta^{0i}\ne 0$, is not unitary, while it is 
unitary if only space-space component $\theta^{ij} \ne 0$ ~\cite{Gomi}. 
Furthermore, a field theory with $\theta^{0i}\ne 0$ and 
$\theta_{\mu\nu}\theta^{\mu\nu}>0$ is still unitary~\cite{CO2}. In this 
case, one can change this to the case with only $\theta^{ij}\ne 0$ by an 
observer Lorentz transformation~\cite{CO2,Carr}. For simplicity, 
here we choose 
\begin{equation}
\label{eq14}
 \theta^{23} \equiv \theta \ne 0,
\end{equation}
and other components vanish identically.  Further, we suppose that the 
background electromagnetic field is of the form
\begin{equation}
\label{eq15}
\bar F_{\mu\nu} = -E U^{01}_{\mu\nu} +BU^{23}_{\mu\nu},
\end{equation}
where $E$ and $B$ are the strengths of electric and magnetic fields, 
respectively, and the 
notation $U^{01}_{\mu\nu}$ has the expression $U^{01}_{\mu\nu}= 
\delta^0_{\mu}\delta^1_{\nu} -\delta^1_{\mu}\delta^0_{\nu}$, the other has a 
similar form. Since one can 
always choose an appropriate coordinate, in which the electric field and
magnetic field are parallel to each other, we therefore assume the
background electromagnetic field to be of the form (\ref{eq15}).  The form
(\ref{eq15}) implies that both the electric and magnetic fields are
parallel to the coordinate axis $x^1$.  Introducing some linearly independent
combinations of momentum components
\begin{equation}
\label{eq16}
l_{\nu} = k^{\mu}U^{01}_{\mu\nu}, \ \  m_{\nu}=k^{\mu}U^{02}_{\mu\nu},\ \
n_{\nu}=k^{\mu}U^{03}_{\mu\nu},
\end{equation}
and another dependent combination $p_{\nu}=k^{\mu}U^{23}_{\mu\nu}$, and using
$l_{\nu}$, $m_{\nu}$ and $n_{\nu}$ to contract (\ref{eq13}), we arrive at
\begin{eqnarray}
\label{eq17}
&& k^2 (a\cdot v)-2 \theta (a\cdot v)[E(l\cdot p)-B(p\cdot p)]
   \nonumber \\
&&~~~~~~~~~  +\theta (k\cdot v)a^{\alpha}U^{23}_{\alpha\beta}(El^{\beta}
   -B p^{\beta}) +\theta [E(l\cdot k)
-B(p\cdot k)]a^{\alpha}v^{\beta}U^{23}_{\alpha\beta}
   \nonumber \\
&&~~~~~~~~~ -\theta (k\cdot v) p^{\alpha}a^{\lambda}(EU^{01}_{\alpha\lambda}
    -BU^{23}_{\alpha\lambda}) +\theta (k\cdot p)a^{\lambda} v^{\delta}(EU^{01}
   _{\lambda\delta}-BU^{23}_{\lambda\delta})=0,
\end{eqnarray}
where $v=l, m, n$, respectively. Now we discuss the propagation of photons 
whose direction is parallel or perpendicular to the one of the background 
electromagnetic field, respectively.   

{\it (a) The Parallel Case}. Namely, the propagation direction of photons
is along $x^1$. In this case, we have 
\begin{equation}
k^{\mu}=(k^0, k^1,0,0), \ \ l^{\mu}=(k^1, k^0,0,0), \ \ m^{\mu}=(0,0,k^0,0),
 \ \ n^{\mu}=(0,0,0,k^0),
\end{equation}
and $p^{\mu}=0$. Substituting these into (\ref{eq17}) and taking $v=l, m, n$,
respectively, it is easy to find that
\begin{equation}
k^2(a\cdot l)=0, \ \  k^2(a\cdot m)=0, \ \  k^2(a\cdot n)=0,
\end{equation}
which implies that the light-cone condition of photons does not change not only
for the longitudinal polarization $(a\cdot l)$, but also for the two 
transverse polarizations $(a\cdot m)$ and ($a\cdot n)$. The light-cone 
condition $k^2=0$ indicates that the velocity of photons does not change in 
this case,
\begin{equation}
\label{eq20}
v_{x^1}=v_{x^2}=v_{x^3}=1.
\end{equation}
 
{\it (b) The Perpendicular Case}. Let us suppose that the propagation of
photon is along $x^3$. In this case, one has 
\begin{equation}
k^{\mu}=(k^0,0,0,k^3), \ \ l^{\mu}=(0, k^0,0,0), \ \ m^{\mu}=(0,0,k^0,0),
\ \ n^{\mu}=(k^3,0,0,k^0),
\end{equation}
and $p^{\mu}=(0,0,-k^3,0)$. Substituting these into (\ref{eq17}),
for the longitudinal polarization $(a\cdot n)$ along $x^3$, we find
\begin{equation}
\label{eq22}
k^2 (a\cdot n) +2\theta B(k^3)^2 (a\cdot n)=0,
\end{equation} 
from which we obtain the velocity of photon along $x^3$
\begin{equation}
\label{eq23}
v_{x^3}=\left| \frac{k^0}{k^3}\right| \approx 1+ \theta B,
\end{equation}
to the leading order of $\theta$.
For the two transverse polarization $(a\cdot l)$ along $x^1$ and $(a\cdot m)$
 along $x^2$, we have 
\begin{equation}
\label{eq24}
(k^2 + 2\theta B(k^3)^2)(a\cdot l)=0, \ \ \ \ 
(k^2 +2\theta B (k^3)^2)(a\cdot m)=0.
\end{equation}
One can see that the light-cone condition is changed, and the velocity
of photons becomes
\begin{equation}
\label{eq25}
v_{x^1}=v_{x^2} \approx 1 + \theta B,
\end{equation}
up to the leading order of $\theta$.

If the propagation direction of photons is along $x^2$, it is easy to show
that this case is the same as the one along $x^3$. That is, the velocity 
of photons is also changed in the same way as Eqs.~(\ref{eq23}) 
and (\ref{eq25}). 

From Eqs.~(\ref{eq20}), (\ref{eq23}) and (\ref{eq25}), we find some interesting
results: in our setup, the background electric field has no  effect on the
propagation of the noncommutative photon; the light-cone condition does not
change for the photons propagating along the direction of background 
magnetic field; but when propagation direction
is perpendicular to the direction of background magnetic field,
the velocity of photons is changed and will be larger than the speed of light
($c=1$) if $\theta B >0$. (For Dp-branes in the magnetic field, one has
$\theta =B^{-1}$ in the decoupling limit~\cite{SW}. But this cannot be 
substituted into (\ref{eq23}) and (\ref{eq25}), since  
$\theta B \ll 1$  is assumed in our approximation.) Thus, the interesting 
question arises of whether this superluminal phenomenon is measurable. If we 
set $\theta \sim (10^4 Gev)^{-2}$ \cite{CST} and $B \sim 1T$, the deviation
from the speed of light is then of the order $10^{-24}$. It seems 
beyond the scope that present experiments can reach.

Some additional remarks are in order. First we note that the shift of 
velocity of photons
is a generic phenomenon in nontrivial vacua (for a review see \cite{Schar}),
for example, in  electromagnetic fields \cite{Adler,Shore}, in  
gravitational fields \cite{DH,Shore,DS1,DS2,Cho,Cai,Dolgov,Shore1}, in 
Casimir-type regions with boundaries \cite{Schar1,Bart}, in finite temperature
backgrounds \cite{Lato} and so on.  In the gravitational fields and the
Casimir-type regions, the velocity of photons can be larger than the 
speed of light and it is always smaller than the speed of light in other 
cases. But our result about the superluminal noncommutative photons is quite
different from those, because the  shift of velocity of photons in 
those nontrivial vacua is essentially  a {\it quantum } phenomenon
because it appears after taking into account one-loop (and/or) two-loop 
corrections to the action of usual QED, while our result [see (\ref{eq23}) 
and (\ref{eq25})] is completely a {\it classical} phenomenon due to the 
noncommutativity of space, since the action (\ref{eq5}) is a classical one 
without any quantum corrections.

Second, a propagating velocity larger than the speed of light is always
associated with a potential violation of causality. For the case of an 
electromagnetic wave traveling in the vacuum between two parallel conducting 
plates (Casimir regions), Ben-Menahem \cite{Ben} showed that the wavefront 
still travels exactly
at the speed of light. So the two-loop effect poses no threat to the causality
in QED. In a flat spacetime, in general the establishment of a causal 
paradox needs at
least two condition: spacelike motion and Poincar\'e invariance. In curved
spaces the Poincar\'e invariance is lost and the principle of equivalence 
replaces it. In the effective action of QED in curved spaces, however, there
are some interaction terms between the electromagnetic field and the
spacetime curvature, which violate the principle of equivalence. Therefore
the appearance of the superluminal photons in curved spaces is possible and
does not necessarily violate the causality. In our case considered in this 
note, although we are discussing the propagation of photons in a flat 
spacetime, the Lorentz symmetry is 
lost due to the noncommutativity of space. As a result, the propagation of 
superluminal photons is allowed in noncommutative spacetimes and does not
threaten the causality.
   
In summary, we have found a very interesting feature of noncommutative 
spacetimes: in which the propagation of photons can be faster than the 
speed of light and it does not necessarily violate the causality.

{\bf Note added:} The same conclusion  has also been obtained 
independently in \cite{GJPP}. Since this note appeared on the net,
I have been informed of many interesting papers, which are related
to the conclusion of this note in some sense. In~\cite{Bak,Has} it has
been shown that solitons in the noncommutative field theories can travel 
faster than the speed of light along the noncommutative directions.
In~\cite{LLT} it has been found that in 
the noncommutative supersymmetric Yang-Mills the low momenta modes have 
superluminal group velocity at finite temperature (this is a one-loop 
effect). However, the propagation of photons in the Born-Infeld theory
is always subluminal~\cite{Gibbons1,Gibbons2}. In addition, the varying
speed of light in noncommutative geometry has been used to solve some 
cosmological problems in the early universe~\cite{Alex}.

\section*{Acknowledgments}
The author would like to thank Prof. N. Ohta for a reading, Profs. G. Gibbons
and R. Jackiw for helpful comments, and especially Prof. R. Jackiw for 
pointing me out an error in the earlier version.  This  work  was supported 
in part by a grant from Chinese Academy of Sciences, and in part by the 
Japan Society for the Promotion of Science and Grants-in-Aid for Scientific 
Research Nos. 99020 and 12640270.



\begin{references}
\bibitem{Minw}S. Minwalla, M.V. Raamasdonk and N. Seiberg, JHEP {\bf 0002},
    020 (2000).
\bibitem{Biga}D. Bigatti and L. Susskind, Phys. Rev. D {\bf 62}, 066004 (2000).
\bibitem{CO1}R.G. Cai and N. Ohta, Phys. Rev. D {\bf 61}, 124012 (2000).
\bibitem{Fisc}W. Fischler, J. Gomis, E. Gorbatov, A. Kashani-Poor,
    S. Paban and P. Pouliot, JHEP {\bf 0005}, 024 (2000).
\bibitem{CST}M. Chaichian, M.M. Sheikh-Jabbari and A. Tureanu,
              Phys. Rev. Lett. {\bf 86}, 2716 (2001).
\bibitem{MS}A. Mazumdar and M.M. Sheikh-Jabbari, hep-th/0012363.
\bibitem{Chu}C.S. Chu, B.R. Greene and G. Shiu, hep-th/0011241.
\bibitem{Haya}M. Hayakawa, Phys. Lett. B {\bf 478}, 394 (2000).
\bibitem{SW}N. Seiberg and E. Witten, JHEP {\bf 0009}, 032 (1999).
\bibitem{GRS}O.J. Ganor, G. Rajesh and S. Sethi, Phys. Rev. D {\bf 62},
       125008 (2000).
\bibitem{Bich}A.A. Bichl, J.M. Grimstrup, L. Popp, M. Schweda and R. 
   Wulkenhaar, hep-th/0102103.
\bibitem{Carr}S.M. Carroll, J.A. Harvey, V.A. Kosteleck\'y, C.D. Lane and
   T. Okamoto, hep-th/0105082.
\bibitem{Adler}S.L. Adler, Ann. Phys. (New York) {\bf 67}, 599 (1971). 
\bibitem{DH}I.T. Drummond and S.J. Hathrell, Phys. Rev. D {\bf 22},
    343 (1980).
\bibitem{Gomi}J. Gomis and T. Mehen, Nucl. Phys. {\bf B591}, 265 (2000).
\bibitem{CO2}R.G. Cai and N. Ohta, JHEP {\bf 0010}, 036 (2000).
\bibitem{Schar}K. Scharnhorst, Annalen Phys. {\bf 7}, 700 (1998). 
\bibitem{Shore}G.M. Shore, Nucl. Phys. {\bf B460}, 379 (1996).
\bibitem{DS1}R.D. Daniels and G.M. Shore, Nucl. Phys. {\bf B425}, 634 (1994).
\bibitem{DS2}R.D. Daniels and G.M. Shore, Phys. Lett. B {\bf 367}, 75 (1996).
\bibitem{Cho}H.T. Cho, Phys. Rev. D {\bf 56}, 6416 (1997).
\bibitem{Cai}R.G. Cai, Nucl. Phys. {\bf B524}, 639 (1998).
\bibitem{Dolgov}A.D. Dolgov and I.D. Novikov, Phys. Lett. B {\bf 442}, 
    82 (1998).
\bibitem{Shore1}G.M. Shore, gr-qc/0012063.
\bibitem{Schar1}K. Scharnhost, Phys. Lett. B {\bf 236}, 354 (1990).
\bibitem{Bart}G. Barton, Phys. Lett. B {\bf 237}, 559 (1990).
\bibitem{Lato}J.I. Latorre, P. Pascual and R. Tarrach, Nucl. 
    Phys. {\bf B437}, 60 (1994).
\bibitem{Ben}S. Ben-Menahem, Phys. Lett. B {\bf 250}, 133 (1990).
\bibitem{GJPP} Z. Guralnik, R. Jackiw, S.Y. Pi and A.  P. Polychronakos,
     hep-th/0106044.
\bibitem{Bak}D. Bak, K. Lee and J.H. Park, Phys. Rev. {\bf D 63}, 125010
            (2001).
\bibitem{Has}A. Hashimoto and N. Itzhaki, hep-th/0012093.
\bibitem{LLT}K. Landsteiner, E. Lopez and M.H.G. Tytgat, JHEP {\bf 0009},
          027 (2000).
\bibitem{Gibbons1}G.W. Gibbons and C.A.R. Herdeiro, Phys. Rev. D {\bf 63},
    064006 (2001).
\bibitem{Gibbons2}G.W. Gibbons, hep-th/0104015. 
\bibitem{Alex}S.H.S. Alexander and J. Magueijo, hep-th/0104093.
\end{references}
\end{document}